\documentclass[fleqn,usenatbib,useAMS]{mnras}

\usepackage{graphicx}	
\usepackage{amsmath}	
\usepackage{amssymb}	
\usepackage{multicol}        
\usepackage{bm}		
\usepackage{pdflscape}	

\usepackage{orcidlink}
\usepackage{subfigure}

\usepackage[T1]{fontenc}
\usepackage{ae,aecompl}

\usepackage{newtxtext,newtxmath}

\title[Short title, max. 45 characters]{MNRAS \LaTeXe\ template -- title goes here}

\title[Cosmology in modified \texorpdfstring{$f(\mathcal{G})$}{} gravity....]{Cosmology in modified \texorpdfstring{$f(\mathcal{G})$}{} gravity: a late time cosmic phenomena}

\author[Santosh V. Lohakare et al.]{Santosh V. Lohakare \orcidlink{0000-0001-5934-3428},$^{1}$\thanks{E-mail: lohakaresv@gmail.com}
Soumyadip Niyogi \orcidlink{0009-0009-7695-7574},$^{2}$\thanks{E-mail: nsoumyadip05@gmail.com}
B. Mishra \orcidlink{0000-0001-5527-3565}$^{1}$\thanks{E-mail: bivu@hyderabad.bits-pilani.ac.in}
\\
$^{1}$ Department of Mathematics, Birla Institute of Technology and Science-Pilani, Hyderabad Campus, Hyderabad 500078, India,\\
$^{2}$Department of Physics, Indian Institute of Science Education and Research Thiruvananthapuram, Thiruvananthapuram-695551, India
}

\date{Accepted XXX. Received YYY; in original form ZZZ}

\pubyear{2024}

\begin{document}
\label{firstpage}
\pagerange{\pageref{firstpage}--\pageref{lastpage}}
\maketitle

\begin{abstract}
    In this work, we present a method for numerically solving the Friedmann equations of modified $f(\mathcal{G})$ gravity in the presence of pressureless matter. This method enables us to predict the redshift behaviour of the Hubble expansion rate. To evaluate the credibility of the model, we applied a Bayesian MCMC technique using late-time cosmic observations to impose limitations on the free parameters of the Gauss-Bonnet model. Our results suggest that the $f(\mathcal{G})$ model can reproduce the low-redshift behaviour of the standard Lambda cold dark matter ($\Lambda$CDM) model, but there are significant differences at high redshifts, leading to the absence of a standard matter-dominated epoch. We also examined the profiles of cosmographic parameters using the model parameter values from the standard range to verify the intermediate epochs. Our analysis shows that the highly promising $f(\mathcal{G})$ model is a feasible candidate for explaining the current epochs. We presented a dynamical system analysis framework to examine the stability of the model. Our study identified critical points depicting various phases of the Universe and explained the evolutionary epochs. We demonstrated that the model effectively captures the evolution of energy components over cosmic time, supporting its validity as an alternate explanation for the observed acceleration of the Universe.
\end{abstract}



\section{Introduction}
    The quest to understand the gravitational force led to the development of general relativity (GR) by \citet{einstein1915}. This theory revolutionized our understanding of gravity by describing it as the curvature of space-time caused by the presence of mass and energy \citep{Misner_1973_book}. Although GR has successfully passed numerous experimental tests \citep{will2014living} and remains the cornerstone of modern gravitational physics, it struggles to fully explain phenomena such as the accelerated expansion of the Universe \citep{Riess_1998_116}, dark matter \citep{Bertone_2005_405}, and dark energy (DE) \citep{COPELAND_2006_15}. In response to these challenges, several modified theories of gravity have been proposed \citep{Clifton_2012_513}.
    
    The most recent observational data, including the Dark Energy Spectroscopic Instrument (DESI) surveys \citep{Adame_2024_DESI_collaboration},  Type Ia Supernovae (SNe Ia) \citep{Riess_1998_116, Perlmutter_1998_517}, Wilkinson microwave anisotropy probe experiment (WMAP) \citep{Spergel_2003_148}, cosmic microwave background (CMB) \citep{Hinshaw_2013_208}, Baryon oscillation spectroscopic survey (BOSS) \citep{Alam_2017_470} and the Baryon Acoustic Oscillations (BAO) data sets \citep{Eisenstein_2005_633} have prompted researchers to consider modifications and expansions to the principles of GR in theories such as \(f(R)\) \citep{Sotiriou_2010_82}, \(f(T)\) \citep{Ferraro_2007_75}, and \(f(Q)\) \citep{Jimenez_2018_98, Heisenberg_2023_1066_Review}. These alternative theories aim to better accommodate and explain the new observational data. One of the most straightforward extensions to Einstein’s gravity is the so-called $f(R)$ gravity, where $f$ is an arbitrary function of the Ricci scalar $R$ \citep{nojiri2007introduction}. Even in this relatively simple case, constructing viable $f(R)$ models consistent with cosmological and local gravity constraints is not straightforward. This complexity arises because $f(R)$ gravity introduces a strong coupling between DE and non-relativistic matter in the Einstein frame \citep{capozziello2008extended}. Extensions of GR, including the Gauss-Bonnet (GB) invariant in the gravitational action, have generated significant interest \citep{Nojiri_2005_631, Li_2007_76, Elizalde_2010_27, Felice_2009_80_063516, Maurya_2021_81, Maurya_2022_925, Nojiri_2017_692, ODINTSOV_2019_938_935, Lohakare_2023_40_CQG, Nojiri_2008_172_81}. One such theory that has garnered significant interest is $f(\mathcal{G})$ gravity. This theory modifies the Einstein-Hilbert action by introducing a function of the GB invariant, denoted $\mathcal{G}$, which is a combination of the Ricci scalar $R$, the Ricci tensor $R_{\mu \nu}$, and the Riemann tensor $R_{\mu \nu \sigma \gamma}$ \citep{Stelle_1978_9, Barth_1983_28, Felice_2009_6, Nojiri_2005_631, nojiri2006modified}. It belongs to an infinite class of curvature invariants known as the Lovelock scalars along with $R$. These do not introduce derivative terms greater than two into the equations of motion for the metric tensor. In four dimensions, the term $\sqrt{-g} \mathcal{G}$ is a total derivative, so the GB term contributes to the equations of motion only when coupled to something else, such as a scalar field $\phi$ with the form $f(\phi) \mathcal{G}$ coupling \citep{Tsujikawa_2002_66, Cartier_2000_2000_035}. A dilaton-graviton mixing term generates this kind of coupling in the low-energy effective action of string theory \citep{Gasperini_2003_373}. 

    The interest in $f(\mathcal{G})$ gravity lies in its potential to explain the observed late-time cosmic acceleration in the Universe. This acceleration could be caused by a gravity modification rather than an unusual source of matter with negative pressure \citep{carroll2004dark}. In recent years, significant research has been conducted into modified gravity to understand the nature of DE \citep{Nojiri_2011_505}. Modified gravity models are particularly attractive because they align more closely with cosmological observations and local gravity experiments than models that rely on exotic matter sources \citep{joyce2015dark}. It is suggested that this theory can pass solar system tests \citep{Davis_2007_0709.4453, Felice_2009_80_063516} and may describe the most exciting features of late-time cosmology, such as the transition from deceleration to acceleration and the current acceleration of the Universe \citep{Nojiri_2005_631, Davis_2007_0709.4453, Felice_2009_80_063516, Cognola_2006_73}.

    In this work, we have explored a subclass of the $f(\mathcal{G})$ model to test its viability as an alternative to the standard cosmological paradigm. We have developed a numerical method to predict the redshift behaviour of the Hubble expansion rate, and our results suggest that the model can reproduce the low-redshift behaviour of the standard Lambda cold dark matter ($\Lambda$CDM) model but has significant differences at high redshifts. The $f(\mathcal{G})$ model is a feasible candidate for explaining the current epochs and effectively captures the evolution of energy components over cosmic time, supporting its validity as an alternative explanation for the observed acceleration of the Universe. We delved into the background cosmological dynamics of the chosen model and evaluated its feasibility using Bayesian analysis supported by Markov Chain Monte Carlo (MCMC) methods applied to late-time cosmic observations, such as Supernovae Ia (Pantheon$^+$) and observational Hubble data (CC sample). We have introduced a dynamical system analysis framework to assess the stability of the model. Our research pinpointed critical points that illustrate different phases of the Universe and elucidated the evolutionary epochs. We have shown that the model effectively represents the changing energy components over cosmic time, supporting its credibility as an alternative explanation for the observed acceleration of the Universe.
    
    This work aims to establish constraints on $f(\mathcal{G})$ cosmology models using CC and Pantheon$^+$ data ses. The paper comprehensively analyses $f(\mathcal{G})$ gravity and uses dynamical system analysis to investigate the stability of the model. The mathematical formalism of $f(\mathcal{G})$ gravity is detailed in Section \ref{SEC II}. In Section \ref{SEC III}, we use the MCMC method to establish correlations between the $f(\mathcal{G})$ gravity model and observational data to determine the best fits for the model parameters $H_0$, $\alpha$, $\beta$, and $m$. Additionally, we present plots of various cosmological parameters such as deceleration, effective equation of state (EoS) $\omega_{\text{eff}} = \frac{p_{\text{eff}}}{\rho_\text{eff}}$, Om diagnostic, and the $r-s$ parameter plot, which are essential for understanding the dynamical behaviour of the Universe under $f(\mathcal{G})$ gravity. Subsequently, in Section \ref{SEC IV}, we construct a dynamical system framework to analyse the critical points of the $f(\mathcal{G})$ gravity model. This analysis is crucial for assessing the stability and viability of the model and its alignment with current cosmological observations. Finally, in Section \ref{SEC V}, we present the conclusions of our results.

\section{\texorpdfstring{$R + f(\mathcal{G})$}{} gravity} \label{SEC II}

    We consider an action that encompasses GR and a functional dependent on the GB term \citep{Nojiri_2005_631, Nojiri_2006_39}:

    \begin{equation}
        S = \int d^4x \sqrt{-g} \left[ \frac{1}{2 \kappa^2} R + f(\mathcal{G}) + \mathcal{L}_m \right],
    \end{equation}
        where $\kappa^2 = 8 \pi G_N = 1$, $G_N$ being the Newtonian constant, $\mathcal{L}_m$ denotes the matter Lagrangian. The GB topological invariant is defined as:

    \begin{equation}
        \mathcal{G} = R^2 - 4 R_{\mu \nu} R^{\mu \nu} + R_{\mu \nu \lambda \sigma} R^{\mu \nu \lambda \sigma}.
    \end{equation}

        By varying the action over $g_{\mu \nu}$, the following field equations are obtained:
    \begin{eqnarray}
        0 &=& \frac{1}{2 \kappa^2} \left( - R^{\mu \nu} + \frac{1}{2} g^{\mu \nu} R \right) + T^{\mu \nu} + \frac{1}{2} g^{\mu \nu} f(\mathcal{G}) - 2 f_\mathcal{G} R R^{\mu \nu} \nonumber \\ & &+ 4 f_\mathcal{G} R^{\mu}_{\rho} R^{\nu \rho} - 2 f_\mathcal{G} R^{\mu\rho\sigma\tau}R^\nu_{\ \rho\sigma\tau} - 4 f_\mathcal{G} R^{\mu\rho\sigma\nu} R_{\rho\sigma} \nonumber \\ & & + 2(\nabla^\mu \nabla^\nu f_\mathcal{G}) R - 2g^{\mu\nu}(\nabla^2f_\mathcal{G})R - 4 (\nabla_{\rho} \nabla^{\mu} f_\mathcal{G}) R^{\nu \rho} \nonumber \\ & & - 4 (\nabla_{\rho} \nabla^{\nu} f_\mathcal{G}) R^{\mu \rho} + 4 (\nabla^2 f_\mathcal{G}) R^{\mu \nu} + 4 g^{\mu \nu} (\nabla_{\rho} \nabla_{\sigma} f_\mathcal{G}) R^{\rho \sigma} \nonumber \\ & & - 4 (\nabla_{\rho} \nabla_{\sigma} f_\mathcal{G}) R^{\mu \rho \nu \sigma},
    \end{eqnarray}

        where we made the notations $f_\mathcal{G} = \frac{df}{d\mathcal{G}}$ and $f_{\mathcal{G}\mathcal{G}} = \frac{d^2 f}{d\mathcal{G}^2}$. We assume a spatially flat FLRW Universe throughout the paper, with a metric given by

    \begin{equation} \label{flrw_metric}
        ds^2 = -dt^2 + a(t)^2 \sum_{i=1}^{3} (dx^i)^2,
    \end{equation}
        where $a(t)$ represents the scale factor at cosmological time $t$ and the GB invariant $\mathcal{G}$ and the Ricci scalar $R$ can be defined as functions of the Hubble parameter as

    \begin{equation}
        \mathcal{G} = 24 \left( \dot{H} H^2 + H^4 \right), \quad R = 6 \left( \dot{H} + 2 H^2 \right).
    \end{equation}
        
        The field equations for the metric \eqref{flrw_metric} yield the FLRW equations in the form:
        \begin{eqnarray}
            3 H^2 &=&\kappa^2 \left(\rho_{\text{m}}+\rho_{\text{r}} + \rho_{\text{DE}}\right) = \kappa^2 \rho_{\text{eff}},  \label{first_field_equation}\\
            \left(2 \dot{H}+3 H^2\right)& =&-\kappa^2 \left(\frac{\rho_{\text{r}}}{3} + p_{\text{DE}}\right) = -\kappa^2 p_{\text{eff}}, \label{second_field_equation}
    \end{eqnarray}
    where $\rho_{\text{m}}$, $\rho_{\text{r}}$ and $\rho_{\text{DE}}$ denotes the matter density, radiation density and DE density, respectively. The over-dot indicates a derivative with respect to cosmic time $t$, and $H \equiv \dot{a}/a$ represents the Hubble parameter. Furthermore, the effective DE density and pressure have been defined as follows:
    \begin{subequations}
    \begin{eqnarray}
        & \rho_{\text{DE}} =\frac{1}{\kappa^2}\Big[ \mathcal{G} f_\mathcal{G} - f(\mathcal{G}) - 24 \dot{\mathcal{G}} H^3 f_{\mathcal{G}\mathcal{G}}\Big], \label{density field eq} \\
        &p_{\text{DE}} = \frac{1}{\kappa^2}\Big[ 8 H^2 \ddot{f}_\mathcal{G} + 16 H \left( \dot{H} + H^2 \right) \dot{f}_\mathcal{G} + f - \mathcal{G} f_\mathcal{G}\Big], \label{pressure field eq}
    \end{eqnarray}
    \end{subequations}

    Without interactions between non-relativistic matter and radiation, these components independently follow their respective conservation laws $\dot{\rho}_{\text{m}}+3 H \rho_{\text{m}}=0$ and $\dot{\rho}_{\text{r}}+4 H \rho_{\text{r}}=0$. From Equations \eqref{density field eq} and \eqref{pressure field eq}, it can be concluded that the DE density and pressure follow the standard evolution equation:
    \begin{eqnarray}
        \dot{\rho}_{\text{DE}}+3 H\left(\rho_{\text{DE}}+p_{\text{DE}}\right)=0.
    \end{eqnarray}
    The parameter for the effective DE EoS can be defined as follows:
    \begin{eqnarray}
        \omega_{\text{DE}} = \frac{p_{\text{DE}}}{\rho_{\text{DE}}}.
    \end{eqnarray}
    In the $\Lambda$CDM limit, as expected, the EoS parameter $\omega_{\text{DE}} \to -1$.

\subsection{Power-Law \texorpdfstring{$f(\mathcal{G})$}{} Model} \label{SEC II a}
        In light of the multiple cosmological data analyses and tests conducted within our solar system, all confirming the principles of GR, one can conclude that any departures from standard GR are expected to be negligible. Consequently, we led to consider the following $f(\mathcal{G})$ functional \citep{Davis_2007_0709.4453, Felice_2009_80_063516},
        \begin{eqnarray}
            f(\mathcal{G}) = \alpha \sqrt{\beta} \left(\frac{\mathcal{G}^2}{\beta^2}\right)^m \, ,
        \end{eqnarray}
        where $\alpha$, $\beta$ and $m$ are positive constants.
        
        In order to determine the theoretical values of the Hubble rate, we can calculate it by solving equation \eqref{density field eq} numerically. When we assume that matter behaves as a pressureless perfect fluid, we can express the matter density as $\rho_\text{m} = 3 H_0^2 \Omega_{\text{m}0} (1+z)^3$, with $z$ representing the cosmological redshift (defined as $\frac{a_0}{a} = 1 + z$, where $a_0$ represents the scale factor at present and $a$ denotes the scale factor when the light emitted) and $\Omega_{\text{m}0}$ representing the current value of the matter density parameter. Thus, for the particular model we are examining, the first Friedmann equation can be expressed as follows:
        \begin{eqnarray} \label{HZ_ode}
            3 H^2 &=& 576^m (2 m-1) \frac{\alpha}{\beta ^{3/2}} H^6 \left(\frac{H^6 \left(H-(z+1) H'\right)^2}{\beta ^2}\right)^{m-1}\nonumber \\& & \times \big(-2 m (z+1)^2 H H''+(z+1) H' \big(2 (3 m-1) H \nonumber \\ & & -(6 m-1) (z+1) H'\big)+H^2\big) + 3 H_0^2 \Omega_{\text{m}0} (1+z)^3 \, ,\nonumber \\
        \end{eqnarray}
        where the prime $(')$ indicates the derivative with respect to $z$.

        Equation \eqref{HZ_ode} represents a second-order differential equation for the function $H(z)$, which can be solved using appropriate boundary conditions. The first initial condition is simply $H(0) = H_0$. For the second initial condition to be determined, we can ensure that, at present, the first derivative of the Hubble parameter is consistent with the predictions of the standard $\Lambda$CDM model, which is characterized by the following expansion law:
        \begin{eqnarray}
            H_{\Lambda \text{CDM}} = H_0 \sqrt{1 - \Omega_{\text{m}0} + \Omega_{\text{m}0} (1+z)^3}  \, ,
        \end{eqnarray}
        After differentiating the above equation with respect to $z$, the second initial condition for equation \eqref{HZ_ode} is obtained as $H'(0) = \frac{3}{2}H_0 \Omega_{\text{m}0}$.

    \section{Observation with numerical solution} \label{SEC III}
        In this part, we will assess the observational feasibility of the model under examination by conducting a Bayesian analysis of the late-time cosmic data. Specifically, we will evaluate the data from the SNe Ia Pantheon$^+$ sample \citep{Brout_2022_938} and the cosmic chronometers (CC) derived from the observational Hubble data compiled in Ref. \citep{Moresco_2022_25}. We have not assumed that the Hubble and Pantheon data sets are correlated. Rather, we will present our results independently for the CC and Pantheon$^+$ data sets. Utilizing these data sets for statistical analysis enables us to obtain reliable results that are not influenced by assumptions of any particular underlying reference model \citep{Agostino_2018_98, Agostino_2019_99, Lohakare_2023_40_CQG}. In the following subsections, we will outline the key characteristics of these measurements and the corresponding likelihood functions.

\subsection{Cosmic Chronometers (CC)} \label{SEC III a}

    \begin{figure*}
    \centering
        \includegraphics[width=88mm]{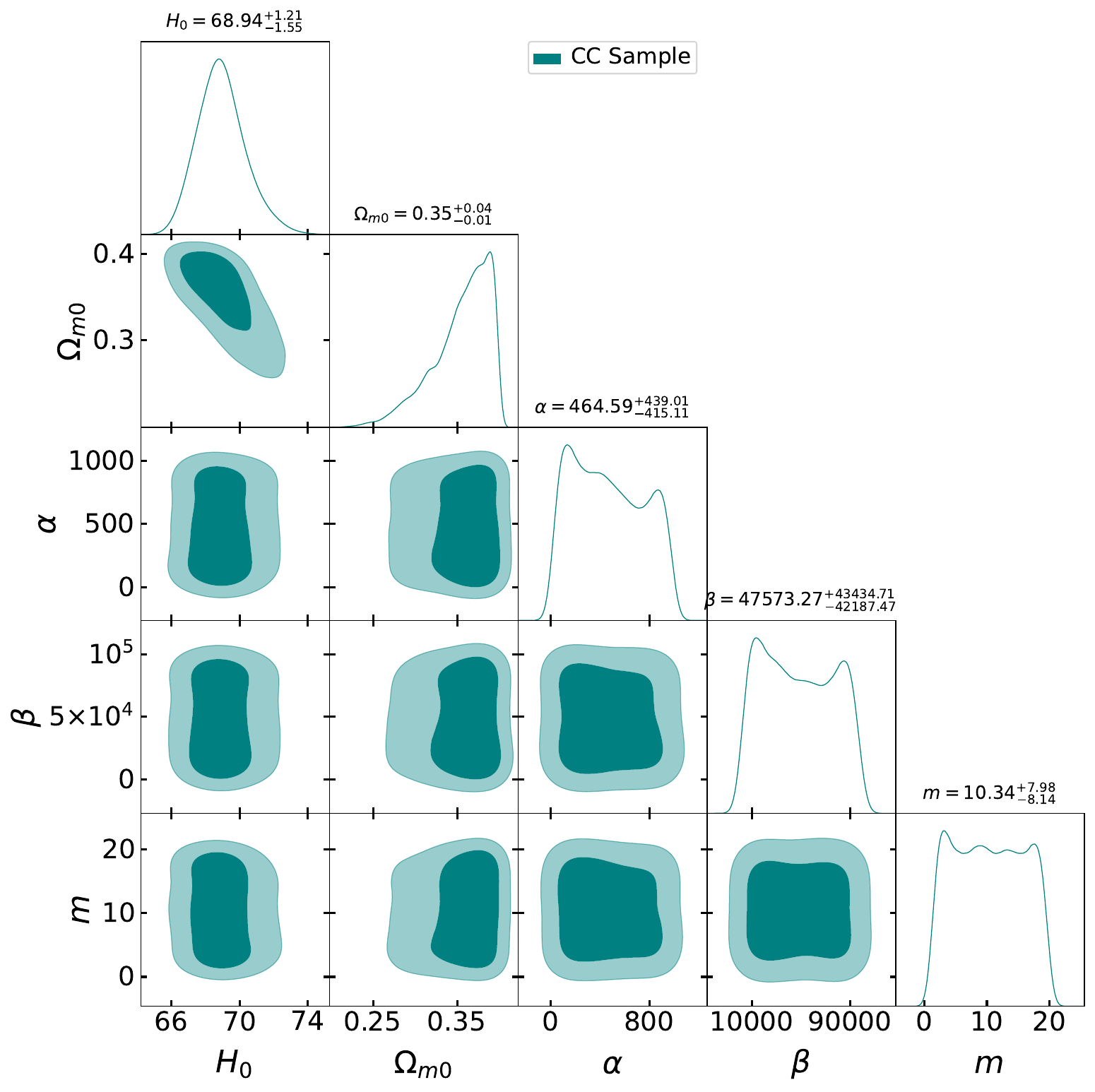}
        \includegraphics[width=88mm]{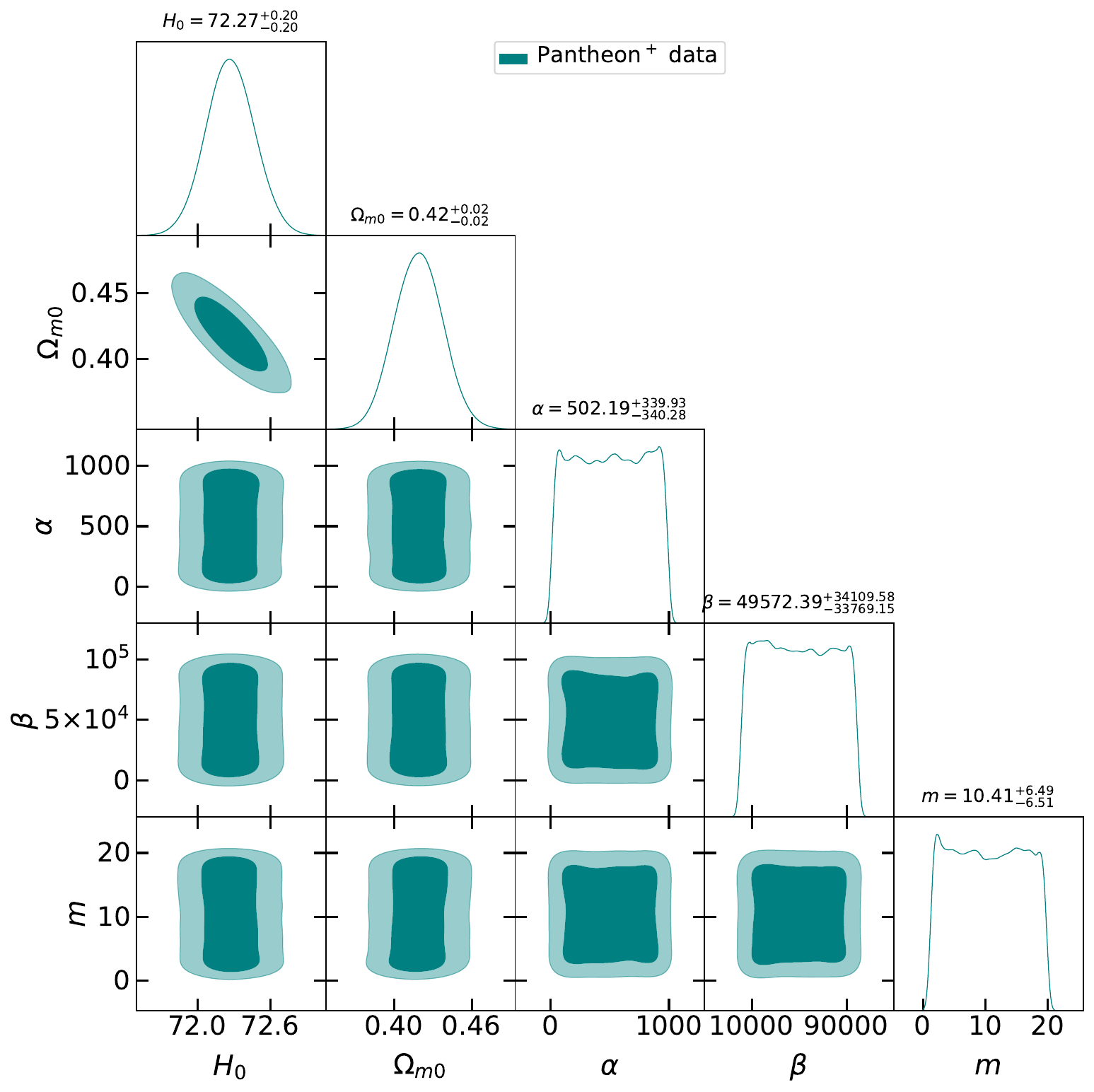}
    \caption{Contour plots show the $1\sigma$ and $2\sigma$ uncertainty regions for the variables $H_0$, $\Omega_{m0}$, $\alpha$, $\beta$, and $m$. These contours are derived from the CC sample (left panel) and the Pantheon$^+$ data (right panel).} 
    \label{FIG1}
    \end{figure*} 

    The Hubble parameter $H(z)$ can be estimated at certain redshifts $z$ using the following formula:
    \begin{equation}
        H(z) = \frac{\dot{a}}{a} = -\frac{1}{1+z}\frac{dz}{dt} \approx -\frac{1}{1+z}\frac{\Delta z}{\Delta t} \, ,  \label{eq: 20}
    \end{equation}

        Here, $\dot{a}$ is the derivative of the scale factor $a$ with respect to time $t$, and $\Delta z$ and $\Delta t$ are the differences in redshift and time, respectively, between two objects. The value of $\Delta z$ can be determined by a spectroscopic survey, while the differential ages $\Delta t$ of passively evolving galaxies can be used to estimate the value of $H(z)$. Compiling such observations can be regarded as a CC sample. We use 32 objects spanning the redshift range $0.07 \leq z \leq 1.965$ \citep{Moresco_2022_25}. For these measurements, one can construct a $\chi^2_{\text{CC}}$ estimator as follows:
    \begin{eqnarray}
        \chi^2_{\text{CC}} = \sum_{i=1}^{32} \frac{[H_{\text{th}}(z_i) - H_{\text{obs}}(z_i)]^2}{\sigma^2_{H}(z_i)} \, , \label{eq: 21}
    \end{eqnarray}

        Here, $H_{\text{obs}}$ and $H_{\text{th}}$ represent the observational and theoretical values of the Hubble parameter, respectively, with $\sigma_H$ being the error in the observational value. Accordingly, we may calculate the $\chi^2_{\text{CC}}$ for both the $\Lambda$CDM model and our specific $f(\mathcal{G})$ model.

\subsection{Supernovae type Ia (SNe Ia)} \label{SEC III b}    
    We will also take into account the Pantheon$^+$ SNe Ia data set, which includes 1701 measurements of the relative luminosity distance of SNe Ia spanning the redshift range of $0.00122 < z < 2.2613$ \citep{Brout_2022_938}. The Pantheon$^+$ compilation consists of distance moduli derived from 1701 light curves of 1550 spectroscopically confirmed SNe Ia within the redshift range of $0.00122 < z < 2.2613$, collected from 18 different surveys. It is worth noting that 77 of the 1701 light curves are associated with galaxies containing Cepheids. The Pantheon$^+$ data set is advantageous in that it can also be used to limit the value of $H_0$ besides the model parameters. To estimate the model parameter from the Pantheon$^+$ samples, we minimize the $\chi^2$ function. To calculate the chi-square $(\chi^2_{\text{SNe}})$ value using the Pantheon compilation of 1701 supernovae data points, we use the following formula:
   \begin{equation}
        \chi^2_{\text{SNe}}= \Delta\mu^T (C_{\text{Sys} + \text{Stat}}^{-1})\Delta\mu \, , \label{eq: 22}
    \end{equation}
        The inverse covariance matrix, $C_{\text{Sys} + \text{Stat}}^{-1}$, associated with the Pantheon$^+$ data set, incorporates both systematic and statistical uncertainties. The term $\Delta \mu$, defined below, signifies the distance residual:
    \begin{equation}
        \Delta\mu = \mu_{\text{th}}(z_i,\theta)-\mu_{\text{obs}}(z_i) \, . \label{eq: 23}
    \end{equation}
        The distance modulus is specifically defined as the difference between an observed apparent magnitude ($m$) of the object and its absolute magnitude ($M$), which quantifies its intrinsic brightness. At a given redshift $z_i$, the distance modulus is expressed as follows:
        \begin{equation}
        \mu_{\text{th}}(z_i,\theta)=5\log_{10}\left(d_L(z,\theta) \right) + 25 = m - M \, , \label{eq: 24}
    \end{equation}
        where $d_L$ denotes the luminosity distance in megaparsecs (Mpc), contingent upon the specific model, which is
        \begin{equation}
        d_L(z,\theta)=\frac{c(1+z)}{H_0}\int_0^z \frac{d\zeta}{E(\zeta)}\, , \label{eq: 25}
    \end{equation}
        where $E(z) = \frac{H(z)}{H_0}$, with $c$ representing the speed of light. Furthermore, the residual distance is indicated by
    \begin{equation}
        \Delta\Bar{\mu}=\begin{cases}
            \mu_k-\mu_k^{cd}, & \text{if $k$ is in Cepheid hosts}\\
            \mu_k-\mu_{\text{th}}(z_k), & \text{otherwise}
        \end{cases} \label{eq: 26}
    \end{equation}
        where $\mu_k^{cd}$ represents the Cepheid host-galaxy distance as determined by SH0ES. This covariance matrix can be integrated with the SNe covariance matrix to form the covariance matrix for the Cepheid host galaxy. Incorporating both statistical and systematic uncertainties from the Pantheon$^+$ data set, the combined covariance matrix is expressed as $C^{\text{SNe}}_{\text{Sys} + \text{Stat}} + C^{cd}_{\text{Sys} + \text{Stat}}$. This formulation defines the $\chi^2$ function for the combined covariance matrix, which is utilized to constrain cosmological models in the analysis:
    \begin{equation}
        \chi^2_{\text{SNe}^+}= \Delta\Bar{\mu} (C^{\text{SNe}}_{\text{Sys}+\text{Stat}}+C^{cd}_{\text{Sys}+\text{Stat}})^{-1}\Delta\Bar{\mu}^T \, . \label{eq: 27}
    \end{equation}
    
    \begin{figure}
    \centering
        \includegraphics[width=75mm]{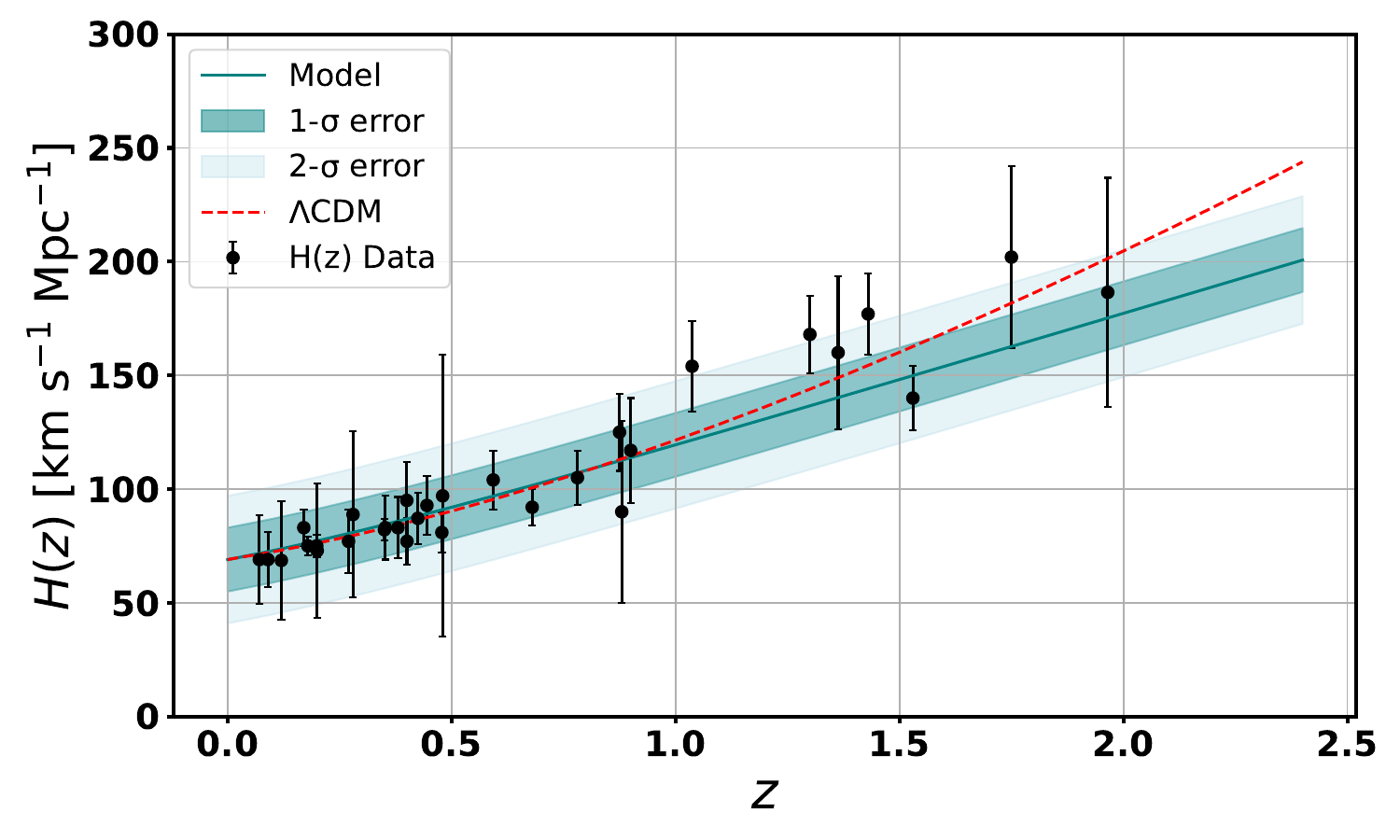}
        \includegraphics[width=75mm]{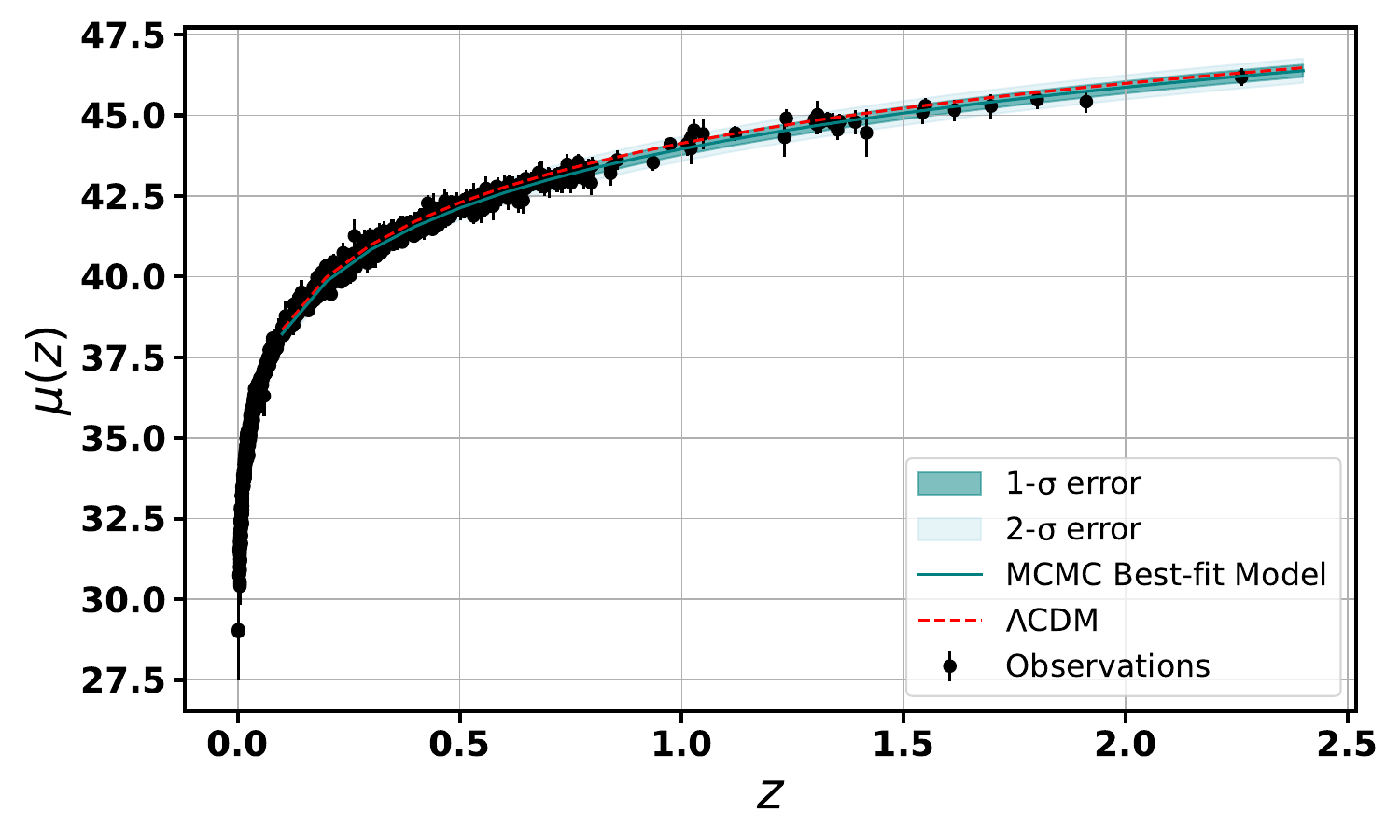}
    \caption{In the upper panel, the black error bars show uncertainty for 32 data points from the CC sample, with the solid teal line representing the model and the broken red line representing $\Lambda$CDM. In the lower panel, the solid teal line represents the distance modulus $\mu(z)$ of the model against redshift $z$, providing a superior fit to the 1701 data points from the Pantheon$^+$ data set with error bars.} 
    \label{FIG2}
    \end{figure} 
    
        We assess the models against the standard $\Lambda$CDM model using the Akaike Information Criterion (AIC) \citep{Akaike_1974_19} and the Bayesian Information Criterion (BIC) \citep{Schwarz_1978_6} in addition to $\chi^2_{\text{min}}$. Both AIC and BIC take into account the model’s goodness of fit as well as its complexity, which is influenced by the number of parameters $(n)$. The AIC is determined as
        \begin{eqnarray}
            \text{AIC} = \chi^2_{\text{min}} + 2 n \, ,
        \end{eqnarray}
        In statistical modelling, a lower AIC value suggests a better fit to the data, taking into consideration the complexity of the model. This penalizes models with more parameters, even if they provide a better fit to the data. Alternatively, the BIC is calculated as
        \begin{eqnarray}
            \text{BIC} = \chi^2_{\text{min}} + n \, \text{ln} \, \mathcal{N},
        \end{eqnarray}
        where $\mathcal{N}$ represents the number of data samples used in the MCMC process. The corrected Akaike Information Criterion ($\text{AIC}_\text{c}$) is defined as
        \begin{eqnarray}
            \text{AIC}_\text{c} = \text{AIC} + \frac{2 n (n+1)}{\mathcal{N}-n-1} \, ,
        \end{eqnarray}
        given that the correction term becomes negligible for large sample sizes $(\mathcal{N} >> n)$, it is not restricted even in such cases. Therefore, it is always advantageous to employ $\text{AIC}_\text{c}$ over the original AIC.
        
        We evaluate the variances in AIC and BIC between the $f(\mathcal{G})$ model and the benchmark model, which is the $\Lambda$CDM model. As a result of this comparison, we can gain a deeper insight into how well each model matches the standard model of cosmology. The differences in AIC and BIC are expressed as $\Delta \text{AIC} = \Delta \chi^2_{\text{min}} + 2 \Delta n$, and $\Delta \text{BIC} = \Delta \chi^2_{\text{min}} + \Delta n \, \text{ln}\, m$, accordingly. A difference in $\text{AIC}_{\text{c}}$ between two competing models can be defined as $\Delta \text{AIC}_\text{c} = \text{AIC}_\text{c {$f(\mathcal{G})$}} - \text{AIC}_\text{c {$\Lambda \text{CDM}$}}$. These measures gauge how each model differs from the benchmark model, with smaller $\Delta$AIC and $\Delta$BIC values suggesting that a model, in conjunction with its selected data set, resembles the $\Lambda$CDM model more closely, indicating superior performance.

    \begin{table*} 
        \centering
    \begin{tabular}{|*{6}{c|}}\hline
        {\centering  \textbf{Data sets}} & $H_0$ & $\Omega_{\text{m}0}$ & $\alpha$ & $\beta$ & $m$ \\ [0.5ex]
    \hline \hline
        {\centering \textbf{CC Sample}} & $68.944^{+1.210}_{-1.551}$ & $0.355^{+0.043}_{-0.014}$ & $464.593^{+439.011}_{-415.110}$ & $47573.273^{+43434.714}_{-42187.475}$ & $10.342^{+7.984}_{-8.144}$  \\
    \hline
        {\centering \textbf{Pantheon}$^+$} & $72.270^{+ 0.200}_{-0.201}$ & $0.420^{+0.020}_{-0.020}$ & $502.193^{+339.937}_{-340.285}$ & $49572.390^{+34109.581}_{-33769.153}$ & $10.412^{+6.491}_{-6.510}$ \\[0.5ex] 
    \hline
    \end{tabular}
        \caption{The table above presents an exploration of the parameters for the MCMC algorithm. It displays the best-fit values for the model parameters, including $H_0$, $\Omega_{\text{m}0}$, $\alpha$, $\beta$, and $m$, derived from the MCMC study using the CC and Pantheon$^+$ data sets.}
        \label{TABLE I a}
    \end{table*}

    \begin{table*}
        \centering
        \begin{tabular}{| *{12}{c|} }
    \hline
        {Data sets}  &\multicolumn{2}{c|}{$\chi^2_{\text{min}}$} &\multicolumn{2}{c|}{AIC} &\multicolumn{2}{c|}{AIC$_\text{c}$} &\multicolumn{2}{c|}{BIC} & {$\Delta \text{AIC}$} & {$\Delta \text{AIC}_\text{c}$} & {$\Delta \text{BIC}$} \\
    \cline{2-9}
        & $f(\mathcal{G})$ & $\Lambda$CDM & $f(\mathcal{G})$ & $\Lambda$CDM & $f(\mathcal{G})$ & $\Lambda$CDM & $f(\mathcal{G})$ & $\Lambda$CDM & & &\\ 
    \hline \hline
        CC sample   & 26.132 & 29.046 & 36.132 & 33.046 & 38.439 & 33.459 & 33.682 & 32.066 & 3.086 & 4.98 & 1.616 \\
    \hline
        Pantheon$^+$  & 1618.774 & 1625.224 & 1628.774 & 1629.224 & 1628.809 & 1629.231 & 1634.924 & 1631.684 & -0.45 & -0.422 & 3.24 \\
    \hline
    \end{tabular}
        \caption{The table provides minimum $\chi^2$ values for the $f(\mathcal{G})$ model, along with their corresponding AIC, AIC$_\text{c}$, and BIC values, and a comparison of AIC, AIC$_\text{c}$, and BIC differences between the model and $\Lambda$CDM.}
        \label{TABLE I b}
    \end{table*}

        The contour plots (see Fig. \ref{FIG1}) display the 1--$\sigma$ and 2--$\sigma$ uncertainty regions for the parameters $H_0$, $\Omega_{\text{m}0}$, $\alpha$, $\beta$, and $m$ using two data sets: the CC data and the Pantheon$^+$ data. These plots, derived using the MCMC method, illustrate the marginalized posterior distributions of parameter pairs, with inner and outer contours representing $68\%$ and $95\%$ confidence levels, respectively.

        The contour plot for the parameters $H_0$ and $\Omega_{\text{m}0}$ exhibits a more elliptical shape compared to the other parameter pairs, which display more square or elongated contours. This elliptical shape indicates that there is a relatively weak correlation between $H_0$ and $\Omega_{\text{m}0}$, suggesting that the data independently constrains these two parameters. This independence implies that the variations in one parameter do not significantly affect the value of the other, leading to a more symmetrical uncertainty region. In contrast, the square or elongated contours seen in other parameter pairs indicate stronger correlations, where changes in one parameter can be offset by adjustments in another to maintain a similar fit to the data. This strong coupling results in less symmetrical and more stretched uncertainty regions, reflecting the interdependence of these parameters within the $f(\mathcal{G})$ gravity model. The best-fit values derived from the MCMC analysis are presented in Table \ref{TABLE I a}. In order to evaluate the efficacy of our MCMC analysis, we calculated the associated AIC, $\text{AIC}_\text{c}$ and BIC values, which are presented in Table \ref{TABLE I b}. Our findings strongly support the assumed $f(\mathcal{G})$ gravity models when analysing the data sets. Moreover, we noted that the $f(\mathcal{G})$ model demonstrates greater precision when applied to the Pantheon$^+$ data sets.

        The upper panel of Fig. \ref{FIG2} shows that the $f(\mathcal{G})$ gravity model fits the observational $H(z)$ data well across the redshift range considered. Both the $f(\mathcal{G})$ gravity and $\Lambda$CDM models follow similar trends, but the $f(\mathcal{G})$ model predicts lower $H(z)$ values at higher redshifts ($z > 1$). This deviation suggests distinct underlying physics due to higher-order curvature terms in the $f(\mathcal{G})$ model, which also accounts for late-time cosmic acceleration without a cosmological constant ($\Lambda$). The lower panel of Fig. \ref{FIG2} illustrates that the $f(\mathcal{G})$ gravity model fits the distance modulus $\mu(z)$ data excellently across the redshift range. Both the $f(\mathcal{G})$ gravity and $\Lambda$CDM models align closely with the observational data, with minimal deviation between them. The strong agreement with observational data supports its viability as a competitive alternative to the $\Lambda$CDM model, with its natural incorporation of higher-order curvature terms making it an attractive option for future cosmological studies. 
        
    \begin{figure*}
        \centering
        \includegraphics[width=160mm]{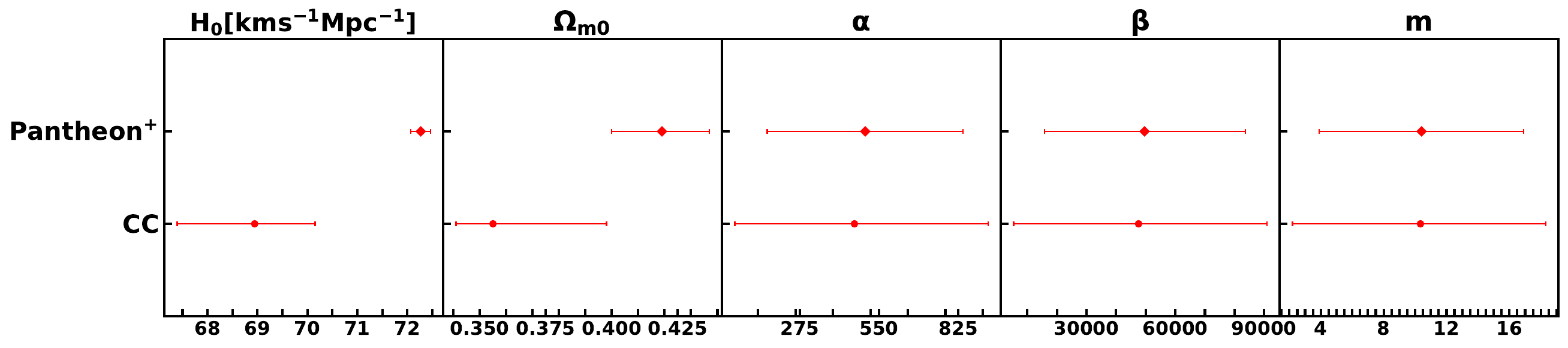}
    \caption{Whisker plot depicting the model parameters $H_0$, $\Omega_{\text{m}0}$, $\alpha$, $\beta$ and $m$, respectively, highlights their discrepancies.} 
    \label{FIG: whisker plot}
    \end{figure*}
    Our analysis reveals a significant discrepancy between the Hubble constant $H_0$ values derived from the CC sample and the Pantheon data sets, as shown in Fig. \ref{FIG: whisker plot}. This whisker plot highlights the ongoing $H_0$ tension in cosmology by presenting the model parameters $H_0$, $\Omega_{\text{m}0}$, $\alpha$, $\beta$, and $m$ along with their $1-\sigma$ confidence intervals.

\subsection{Cosmological Parameter Evolution}\label{SEC III c}
        We analyse the evolution of crucial cosmological parameters, including the effective EoS, statefinder, and Om diagnostic parameters, by imposing constraints on model parameters using various observational data. We present a fully general expression for the deceleration parameter, $q = -\ddot{a}/aH^2$, as follows:
        \begin{eqnarray}
            q(z) = -1 + \frac{H'(z)}{H(z)} (1+z)\, .
        \end{eqnarray}

        \begin{figure}
        \centering
        \includegraphics[width=80mm]{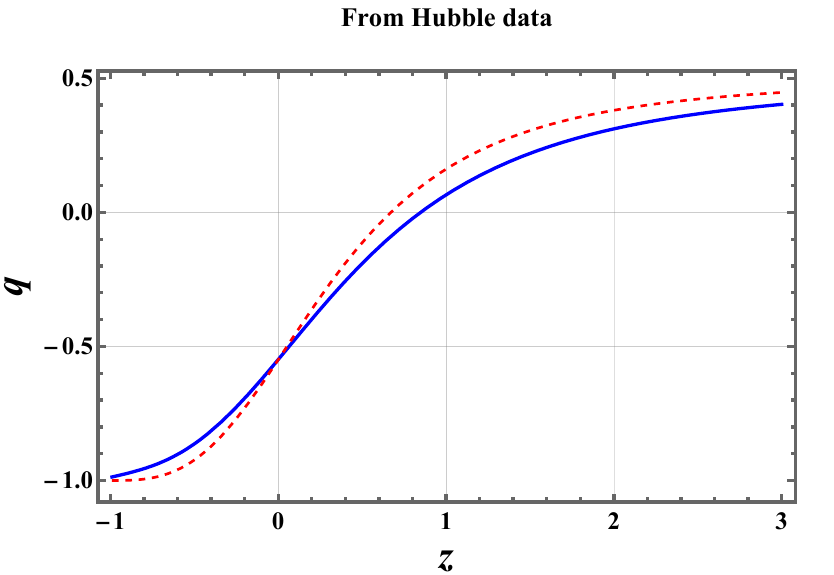}
        \includegraphics[width=80mm]{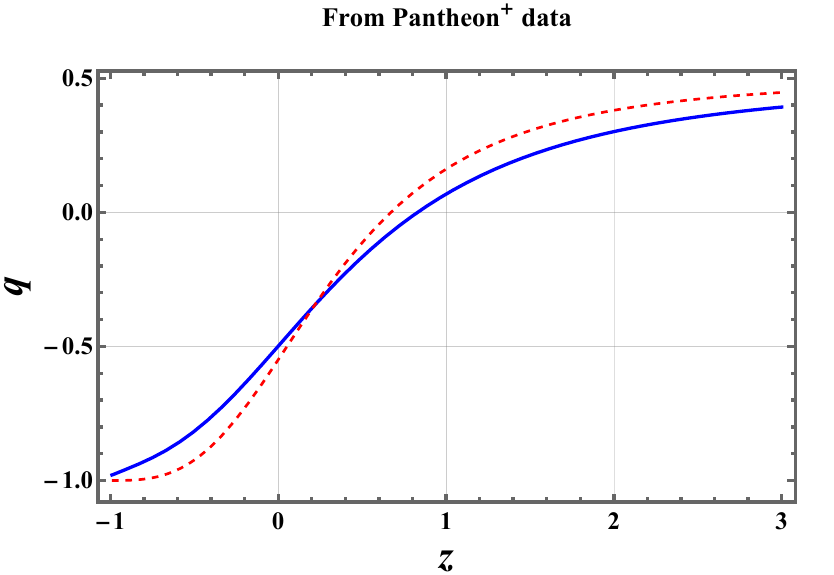}
        \caption{Graphical representation of the deceleration parameter versus redshift using the constrained coefficients from Fig. \ref{FIG1}. The thick line represents the behaviour of the deceleration parameter for the $f(\mathcal{G})$ model, while the dashed line shows the deceleration parameter for the $\Lambda$CDM model.}
        \label{FIG3}
        \end{figure}

        Fig. \ref{FIG3} presents the deceleration parameter as a function of redshift $z$ for the $f(\mathcal{G})$ gravity model, derived from both Hubble data (upper panel) and Pantheon$^+$ data (lower panel). Fig. \ref{FIG3} demonstrate that the restricted values of model parameters derived from the analysed CC and Pantheon$^+$ data sets indicate a transition of $q$ from positive (indicating early deceleration) in the past to negative (indicating current acceleration) in the present. The present value of deceleration parameter $q_0$ is measured to be $-0.527$ and $-0.499$ for the CC and Pantheon$^+$ data, respectively, at the current cosmic epoch, which aligns relatively well with the range of $q_0 = -0.528^{+0.092}_{-0.088}$ determined by recent observations \citep{Christine_2014_89}. Recent observations are consistent with this deceleration parameter, and the resulting model indicates a smooth transition from deceleration to acceleration at $z_t = 0.84$ and $z_t = 0.82$ for the CC and Pantheon$^+$ data sets, respectively. The derived transition redshift $z_t$ aligns with current constraints based on 11 $H(z)$ observations reported by \citet{Busca_2013_552} for redshifts $0.2 \leq z \leq 2.3$, $z_t = 0.74 \pm 0.5$ from \citet{Farooq_2013_766}, $z_t = 0.7679^{+0.1831}_{-0.1829}$ by \citet{Capozziello_2014_90_044016}, and $z_t = 0.60^{+0.21}_{-0.12}$ by \citet{Yang_2020_2020_059}. The consistency between the curves from Hubble and Pantheon$^+$ data underscores the robustness of the $f(\mathcal{G})$ model in capturing the expansion dynamics of the Universe.
    
    \begin{figure}
        \centering
        \includegraphics[width=80mm]{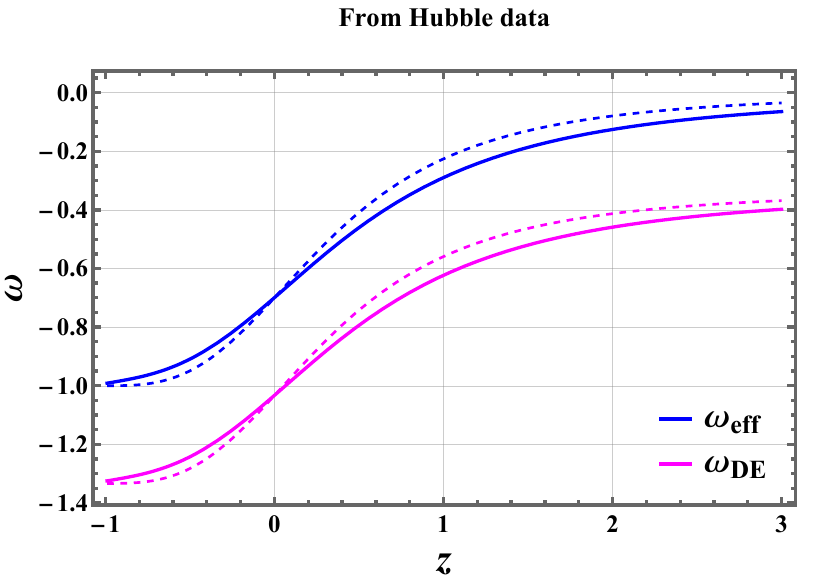}
        \includegraphics[width=80mm]{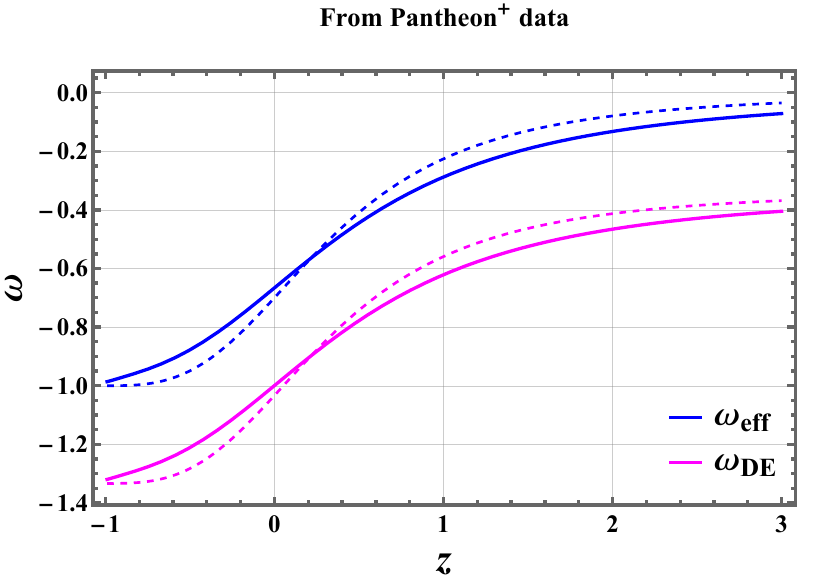}
    \caption{Graphical representation of the EoS parameter versus redshift using the constrained coefficients from Fig. \ref{FIG1}. The thick line represents the behaviour of the EoS parameter for the $f(\mathcal{G})$ model, while the dashed line shows the EoS parameter for the $\Lambda$CDM model.}
    \label{FIG4}
    \end{figure}

    The deceleration parameter is one of the key factors that characterize the behaviour of the Universe, determining whether it continuously decelerates, accelerates, or undergoes multiple phases of transition. Similarly, energy sources influence the evolution of the Universe through the EoS parameter, defined as $\omega_{\text{DE}}$ is shown in Fig. \ref{FIG4}. By calculating the energy density and pressure of DE, as depicted in Fig. \ref{FIG4}, we can observe the variations in the effective EoS of DE relative to the redshift variable. The current EoS values for DE, $\omega_{\text{DE}}(z = 0)$, are obtained as $-1.018$, $-0.999$ for the CC, Pantheon$^+$, data sets, respectively. These values indicate phantom behaviour (at $z < 0$) and a trend towards approximately $-1.32$ at late times. The present values of $\omega_{\text{eff}}$ are $-0.684$, $-0.666$ for the CC and Pantheon$^+$ data sets, respectively. Various cosmological studies have also constrained the EoS parameter, including the Supernovae Cosmology Project $\omega_{\text{DE}}=-1.035^{+0.055}_{-0.059}$ \citep{Amanullah_2010_716}, Planck 2018 $\omega_{\text{DE}}=-1.03\pm 0.03$ \citep{Aghanim_2020_641}, and WAMP+CMB $\omega_{\text{DE}}=-1.079^{+0.090}_{-0.089}$ \citep{Hinshaw_2013_208}.

    Fig. \ref{FIG4} shows the effective EoS, $\omega_{\text{eff}}$ as a function of redshift $z$ for the $f(\mathcal{G})$ gravity model. At low redshifts ($z \approx 0$), $\omega_{\text{eff}}$ is close to $-1$, indicating DE dominance and accelerated expansion. As $z$ increases, $\omega_{\text{eff}}$ transitions from negative values to less negative values, reflecting a shift from a deceleration phase in the early Universe to an acceleration-dominated phase in the current epoch. The consistent behaviour of $\omega_{\text{DE}}$ across both data sets reinforces the capability of the $f(\mathcal{G})$ model to describe the expansion history of the Universe. This transition aligns with theoretical expectations of the $f(\mathcal{G})$ model, which incorporates higher-order curvature terms to account for cosmic dynamics without a cosmological constant. The observed $\omega_{\text{DE}}$ behaviour highlights the effectiveness of the model, supporting its viability as an alternative to the $\Lambda$CDM model. \\
    
    \begin{figure}
        \centering
        \includegraphics[width=80mm]{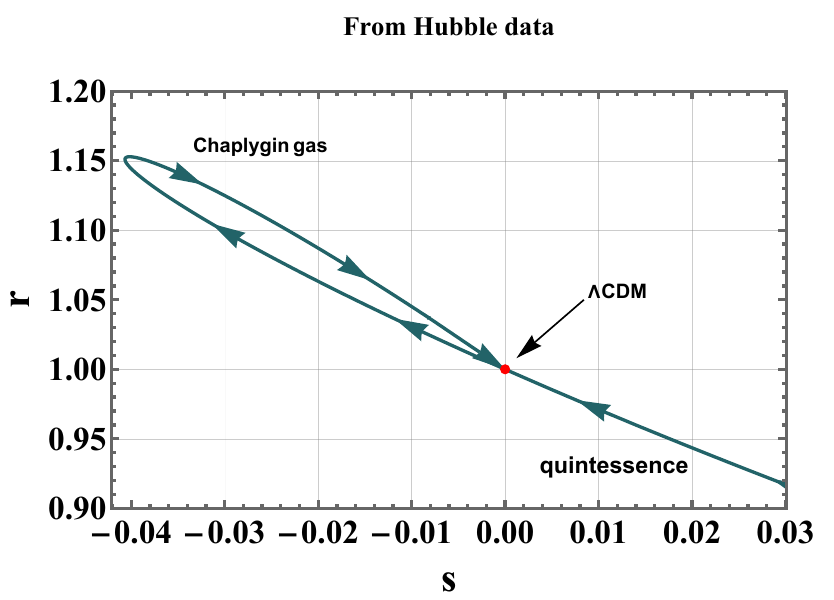}
        \includegraphics[width=80mm]{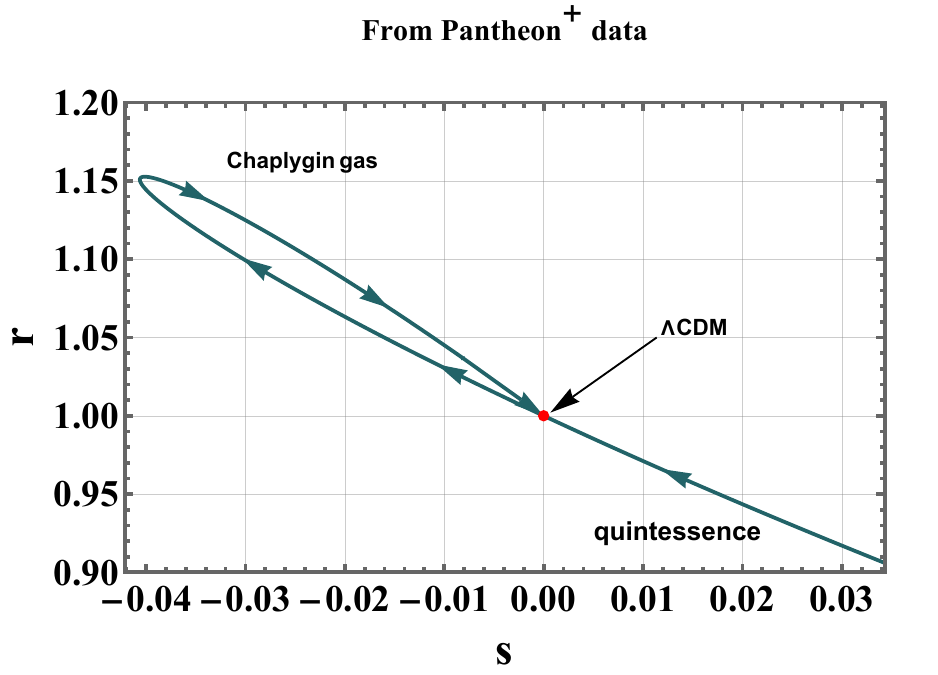}
    \caption{A plot showing the evolution of the given cosmological model in the $r - s$ plane using the constrained coefficients from Fig. \ref{FIG1}.} 
    \label{FIG5}
    \end{figure}

        The statefinder diagnostic proposed by V. Sahni \citep{Sahni_2003_77_201} provides a geometric method for discerning different DE models using statefinder parameters.
        \begin{eqnarray}
            r = \frac{\dot{\ddot{a}}}{a H^3}, \,\,\,\,\,\,\,\,\,\, s = \frac{r - 1}{3 (q-\frac{1}{2})},
        \end{eqnarray}
        The conditions $(r < 1, s > 0)$ correspond to the quintessence of DE, while the domain $(r > 1, s < 0)$ represents the phantom scenario. Additionally, the state $(r = 1, s = 0)$ reproduces the standard $\Lambda$CDM model.

        Fig. \ref{FIG5} illustrates the $r-s$ parameter plot for the $f(\mathcal{G})$ gravity model. The trajectory in the $r-s$ plane highlights the evolutionary track of the expansion of the Universe. The $f(\mathcal{G})$ model passes through the region corresponding to the $\Lambda$CDM model, indicated by the red point. At lower values of $s$, the model aligns with quintessence characteristics, suggesting a dynamical DE component with $\omega > -1$. As $s$ increases, the trajectory moves towards regions associated with Chaplygin gas models, indicating a unified dark matter and DE scenario. The smooth transition observed in the $r-s$ parameter space demonstrates the flexibility of the $f(\mathcal{G})$ model in describing different cosmological behaviours. This capability allows the model to account for various dynamics, from quintessence-like to Chaplygin gas-like, providing a comprehensive description of the expansion of the Universe. 

        The $\text{Om}(z)$ diagnostic is a simple testing method that depends only on the first-order derivative of the cosmic scale factor. In particular, in DE theories, the $\text{Om}(z)$ parameter is followed as an additional effective diagnostic tool \citep{Sahni_2008_78_103502, Sahni_2003_77_201} and is defined as
    \begin{equation} \label{Omz}
        \text{Om}(z)=\frac{\frac{H^2(z)}{H^2_0}-1}{(1+z)^3-1}.
    \end{equation}
        The diagnostic for the two-point difference is
    \begin{equation}
        \text{Om}(z_1, z_2)= \text{Om}(z_1) - \text{Om}(z_2)\, ,
    \end{equation}
        Alternatively stated for simplification, when $\text{Om}(z_1, z_2) > 0$, it indicates quintessence, whereas when $\text{Om}(z_1,z_2) < 0$, it signifies phantom behaviour, where $(z_1 < z_2)$. The $\text{Om}(z)$ diagnostic in the $\Lambda$CDM model serves as a null test, as noted in \citep{Sahni_2008_78_103502}, and its sensitivity to the EoS parameter was further explored in subsequent data as seen in \citep{Qi_2018_18_066, Zheng_2016_825_17, Ding_2015_803_L22}. If $\text{Om}(z)$ remains constant for the redshift, the DE concept would be a cosmological constant. The DE concept will form a cosmological constant if $\text{Om}(z)$ is constant for the redshift. The slope of $\text{Om}(z)$, which is positive for the emerging $\text{Om}(z)$ and denotes phantom phase $(\omega<-1)$ and negative for quintessence region $(\omega > -1)$ also identifies the DE models.
    \begin{figure}
        \centering
        \includegraphics[width=80mm]{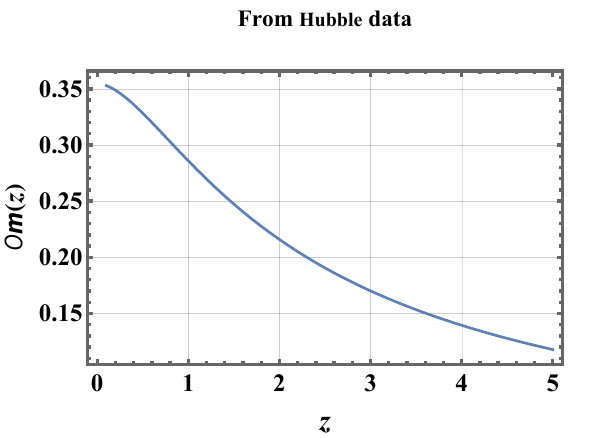}
        \includegraphics[width=80mm]{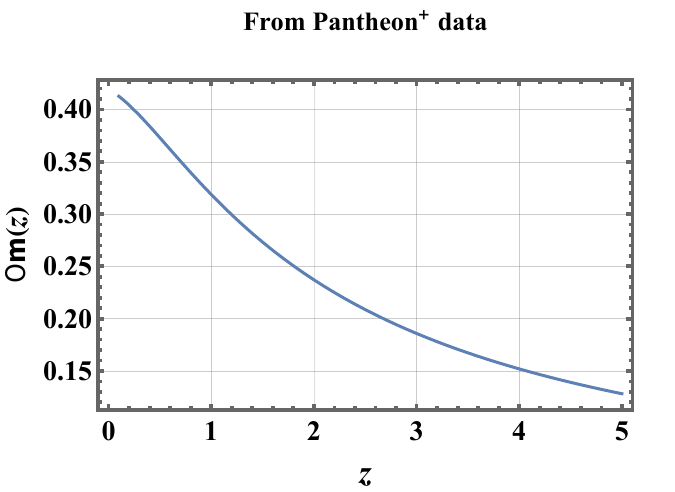}
    \caption{$\text{Om}(z)$ diagnostic parameter profile for the cosmological model using the constrained coefficients from Fig. \ref{FIG1}.} 
    \label{FIG6}
    \end{figure}

        The graph in Fig. \ref{FIG6} shows the reconstructed $\text{Om}(z)$ parameter based on the best-fitting data, plotted against redshift. It illustrates a decreasing trend in the $\text{Om}(z)$ parameter as redshift increases. At higher redshifts ($z > 1$), the Om diagnostic curve shows a significant decline, indicating the deviation of the $f(\mathcal{G})$ gravity model from the $\Lambda$CDM model. This behaviour suggests that the higher-order curvature terms in the $f(\mathcal{G})$ model influence the cosmic dynamics differently compared to the standard cosmological model. The shape of the Om diagnostic curve for the $f(\mathcal{G})$ gravity model highlights its potential to account for cosmic acceleration through modifications to gravity. This diagnostic tool effectively illustrates the differences between the $f(\mathcal{G})$ model and the $\Lambda$CDM model, reinforcing the former’s viability as an alternative explanation for the observed acceleration of the Universe.

\section{Stability Assessment via Dynamical Systems} \label{SEC IV}
        The techniques of dynamical systems are helpful in examining the overall long-term behaviour of a specific cosmological model \citep{Carloni_2005_22, Amendola_2007_75, Ivanov_2012_18, Carloni_2015_92, Duchaniya_2023_83, Duchaniya_2022_82} and can assist in circumventing the challenge of solving non-linear cosmological equations. These techniques characterize the universal dynamics by studying the nearby long-term behaviour of critical points of the system and connecting them to the primary cosmological eras \citep{Paliathanasis_2017_08}.
    
        We used the autonomous dynamical system method to study the problem due to the complicated form of the equation \eqref{density field eq}. For a general $f(\mathcal{G})$ model, it will be helpful to introduce the following variables:
    \begin{align}
        X = H^2 f_{\mathcal{G}} \,,\quad Y = H \dot{f}_{\mathcal{G}} \,,\quad Z=\frac{\dot{H}}{H^2} \,,\quad 
        W=-\frac{f}{6H^2}\,,\quad V=\frac{\kappa^2 \rho_r}{3H^2}\, .\label{dynamical variables}
    \end{align}
    alongside the density parameters:
    \begin{align}
        \Omega_{\text{m}}=\frac{\kappa^2 \rho_{\text{m}}}{3H^2}, \quad \Omega_{\text{r}}=V=\frac{\kappa^2 \rho_{\text{r}}}{3H^2}, \quad \Omega_{\text{DE}}=\frac{\kappa^2 \rho_{\text{DE}}}{3H^2},
    \end{align}
        with the constraint
    \begin{align}   \Omega_{\text{m}}+\Omega_{\text{r}}+\Omega_{\text{DE}}=1.
    \end{align}
        In terms of dynamical variables, we have:
    \begin{align} \label{constraint eq}
        \Omega_{\text{m}}+\Omega_{\text{r}} + 8 X Z + 8 X + 2 W - 8 Y=1,
    \end{align}
        and
    \begin{align}
        \Omega_{\text{DE}} = 8 X Z + 8 X + 2 W - 8 Y.
    \end{align}
        In order to study the time-dependent behaviour of the dynamical system, it is necessary to establish a dimensionless time parameter. In this study, we choose to use a time parameter expressed as the number of e-folds $N \equiv \text{ln}\, a/a_0$, where $a_0$ is a constant with the same units as $a$, and is typically set as $a_0 \equiv 1$. The evolution of each variable is then determined by its derivative with respect to $N$, which is expressed as follows:
    \begin{subequations}\label{generaldynamicalsystem}
    \begin{eqnarray}
        \frac{dX}{dN}& = & 2XZ + Y\,,\\
        \frac{dY}{dN}& = & Y Z + (3 X - 2 Y) (Z + 1) + \frac{3 W}{4} - \frac{Z}{4} - \frac{3}{8} - \frac{V}{8}\,,\nonumber\\ \\
        \frac{dZ}{dN}& = & \lambda - 2 Z^2\,, \\
        \frac{dW}{dN}& = & - 4 X (4 Z + 2 Z^2 + \lambda)-2 W Z\,, \\
        \frac{dV}{dN}& = & -2 V (2 + Z) \,.
    \end{eqnarray}
    \end{subequations}

    \begin{table*}
    \centering
    \begin{tabular}{|*{8}{c|}}\hline 
        {\textbf{Critical Point}} & $\textbf{X}$ & $\textbf{Y}$ & $\textbf{Z}$ & $\textbf{V}$ & \textbf{Existence} & $\omega_{\text{eff}}$ & $q$\\ [0.5ex]\hline \hline 
        {$\mathcal{P}_1= (x_1,y_1,z_1,v_1)$} & $x_1$ & $4 x_1$ & $-2$ & $1 + 24 x_1$ & \begin{tabular}{@{}c@{}}$x_{1} \neq 0$,\,\, $m = \frac{1}{4}$ \end{tabular} & $\frac{1}{3}$ & $1$ \\
    \hline
        {$\mathcal{P}_2= (x_2,y_2,z_2,v_2)$} & $x_{2}$ & $\frac{1}{4} (-1-8 x_2)$ & $1+ \frac{1}{8 x_2}$ & $0$ &  \begin{tabular}{@{}c@{}}$x_{2} \neq 0$,\,\, $m = \frac{1}{4}$ \end{tabular}  & $0$ & $\frac{1}{2}$ \\
    \hline
        {$\mathcal{P}_3 = (x_3,y_3,z_3,v_4)$} & $\frac{-1}{16}$ & $\frac{-1}{8}$ & $-1$ & $0$ &  \begin{tabular}{@{}c@{}}$ 2 m^2 - m \neq 0$ \end{tabular} & $- \frac{1}{3}$ & $0$ \\
    \hline
        {$\mathcal{P}_4 = (x_4,y_4,z_4,v_4)$} & $\frac{m}{-4+8 m}$ & $0$ & $0$ & $0$ & \begin{tabular}{@{}c@{}}$(-1+2 m) (-1+ 4 m) \neq 0$, $m \neq 0$ \end{tabular} & $-1$ & $-1$ \\
    \hline
    \end{tabular}
        \caption{The critical points and physical characteristics of the system.}
    \label{TABLE-II}
    \end{table*}   
    
    \begin{table}
    \centering 
    \begin{tabular}{|*{8}{c|}}\hline 
        {\textbf{Critical Point}} & $\Omega_{\text{m}}$ & $\Omega_{\text{r}}$ & $\Omega_{\text{DE}}$ & $\text{Acceleration}$\\ [0.5ex]\hline \hline 
        {$\mathcal{P}_1$} & $0$ & $1+ 24 x_1$ & $-24 x_1$ & Never\\
    \hline
        {$\mathcal{P}_2$} & $1 + 20 x_{2}$ & $0$ & $-20 x_2$ & Never\\
    \hline
        {$\mathcal{P}_3$} & $0$ & $0$ & $1$ & Never\\
    \hline
        {$\mathcal{P}_4$} & $0$ & $0$ & $1$ & Always\\
    \hline
    \end{tabular}
        \caption{The density parameters associated with the critical point.}
    \label{TABLE-III}
    \end{table}

        \begin{table*}
    \centering
    \begin{tabular}{|*{8}{c|}}\hline 
        {\textbf{Critical Point}} & $\textbf{Eigenvalues}$ & $\textbf{Stability}$ \\ [0.5ex]\hline \hline 
        {$\mathcal{P}_1$} & {$\left\{0, 1, \frac{-x_1+\sqrt{-x_1 (2+47 x_1)}}{2 x_1}, \frac{-x_1-\sqrt{-x_1 (2+47 x_1)}}{2 x_1} \right\}$} & Unstable \\
    \hline
        {$\mathcal{P}_2$} & {$\left\{0, -1, \frac{- 3 x_2+\sqrt{-4 x_2 - 71 {x_2}^2)}}{4 x_2}, \frac{- 3 x_2-\sqrt{-4 x_2 - 71 {x_2}^2)}}{4 x_2} \right\}$} & Stable for $-\frac{4}{71} \leq x_2 < -\frac{1}{20}$ \\
    \hline
        {$\mathcal{P}_3$} &  {$\left\{-2, -2, -1, 4+\frac{2}{-1+2 m} \right\}$} & Stable for $\frac{1}{4} < m <\frac{1}{2}$ \\
    \hline
        {$\mathcal{P}_4$} & {$\left\{-4, -3, \frac{3 m - 6 m^2 -\sqrt{(1-2 m)^2 (25 m -4) m}}{2 m (2 m - 1)}, \frac{3 m - 6 m^2 + \sqrt{(1-2 m)^2 (25 m -4) m}}{2 m (2 m - 1)} \right\}$} & Stable for $\frac{4}{25} \leq m < \frac{1}{4}$ \\
    \hline
    \end{tabular}
    \caption{Eigenvalues and stability regime.}
    \label{TABLE-IV}
    \end{table*}

    Taking into account that $f(\mathcal{G}) = \alpha \sqrt{\beta} \left(\frac{\mathcal{G}^2}{\beta^2}\right)^m$, from equation \eqref{dynamical variables} we get
    \begin{eqnarray} \label{dependent_w}
        W = \frac{2 X (1+Z)}{-m}
    \end{eqnarray}

        In order to obtain an expression for $\lambda = \frac{\ddot{H}}{H^3}$, we can use equation \eqref{dependent_w}:
    \begin{eqnarray}
        \lambda = -\frac{m W (4 X Z + Y) + 8 m X^2 Z (Z + 2) + 4 X^2 Z^2}{2 (2 m-1) X^2} \nonumber \\
    \end{eqnarray}
        As per the given relations and the constraint \eqref{constraint eq} and dependency relation \eqref{dependent_w}, we can remove the equations for $W$ from our autonomous system, leaving us with only a set of four equations:
    \begin{subequations}\label{Autonomous DS}
        \begin{eqnarray}
            \frac{dX}{dN}& = & 2XZ + Y\,,\\
            \frac{dY}{dN}& = & -\frac{3 X Z}{2 m}-\frac{3 X}{2 m}-\frac{V}{8} + 3 X Z + 3 X - Y Z - 2  Y - \frac{Z}{4}-\frac{3}{8}\,,\nonumber\\ \\
            \frac{dZ}{dN}& = & \frac{(Z + 1) \big(Y + 4 (1-2 m) X Z \big)}{(2 m-1) X}\,, \\
            \frac{dV}{dN}& = & -2 V (2 + Z) \,.
        \end{eqnarray}
    \end{subequations}

    \begin{figure*}
    \centering
    \subfigure[]
        {\includegraphics[width=60mm]{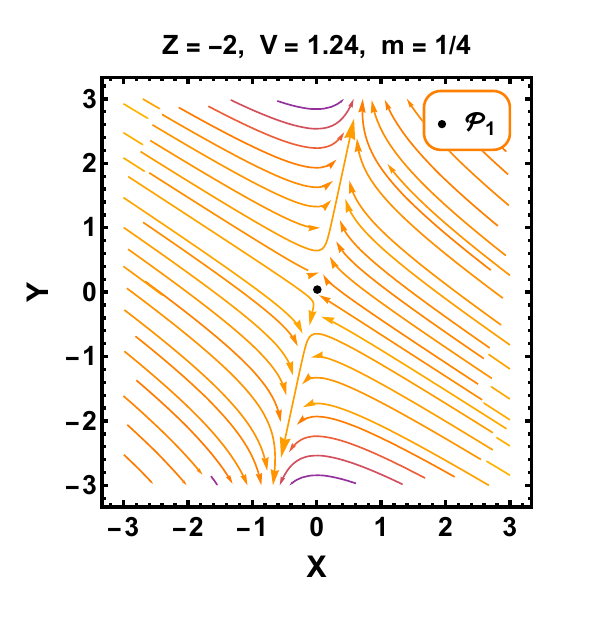} \label{portrait_P1}} 
    \subfigure[]
        {\includegraphics[width=60mm]{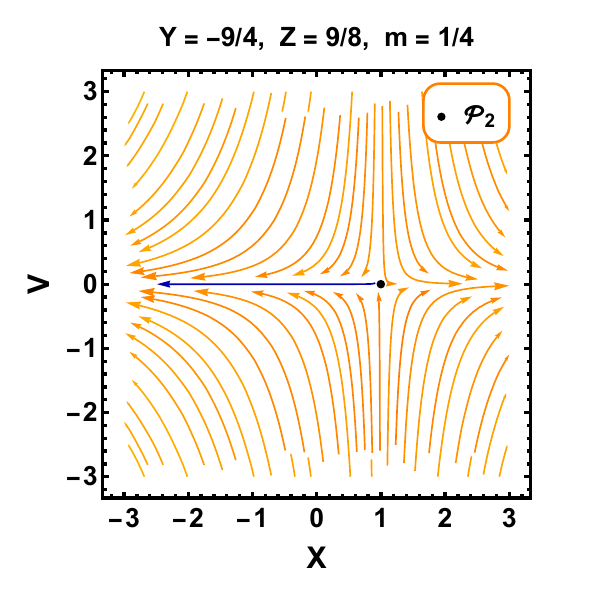} \label{portrait_P2}} \\
    \subfigure[]
        {\includegraphics[width=60mm]{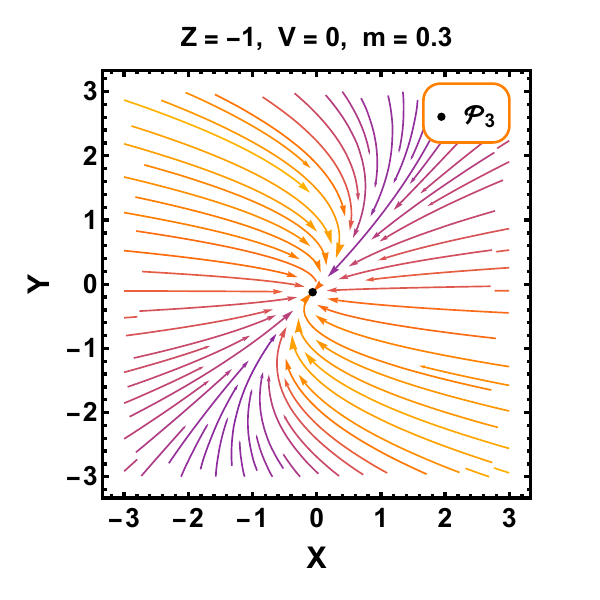} \label{portrait_P4}}
    \subfigure[]
        {\includegraphics[width=60mm]{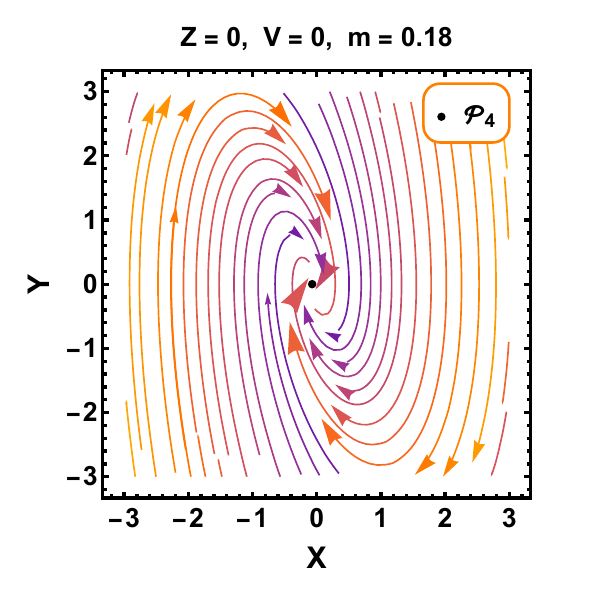} \label{portrait_P5}}
    \caption{Two-dimensional phase portrait for the dynamical system.} 
    \label{FIG7}
    \end{figure*}

        The two-dimensional phase portraits shown in Fig. \ref{FIG7} depict the dynamics of the system for \( m = \frac{1}{4}, 0.3, 0.18 \) by mapping the trajectories on to the \( XY \) and \( XV \) planes. These visualizations provide insights into the stability and nature of the critical points \( \mathcal{P}_1 \) to \( \mathcal{P}_4 \).
    \begin{itemize}
        \item {\bf Critical point \( \mathcal{P}_1 \):} The critical point \( \mathcal{P}_1 \) is identified on the \( XY \) plane in Fig. \ref{portrait_P1}. The phase portrait indicates an unstable node. The trajectories are seen diverging away from the critical point in all directions, confirming the instability of \( \mathcal{P}_1 \) as suggested by its eigenvalues. This divergence implies that small perturbations will cause the system to evolve away from \( \mathcal{P}_1 \), never allowing it to settle into a steady state.
        
    \item {\bf Critical point \( \mathcal{P}_2 \):}
        The critical point \( \mathcal{P}_2 \) is depicted in the \( XV \) plane in Fig. \ref{portrait_P2}. The critical point $\mathcal{P}_2$ corresponds to a non-standard CDM-dominated epoch in which the density of DE is negligible ($\Omega_{\text{DE}} = -20 x_2$). When $x_2 = 0$, this critical point reflects a standard cold dark matter-dominated era. The phase portrait shows trajectories approaching \( \mathcal{P}_2 \) along specific paths, forming a saddle point structure. This behaviour indicates that \( \mathcal{P}_2 \) is a saddle point, with some trajectories being attracted towards it along stable manifolds and repelled away along unstable manifolds. This behaviour is consistent with the mixed stability eigenvalues obtained for \( \mathcal{P}_2 \).

    \item {\bf Critical point \( \mathcal{P}_3 \):}
        The critical point \( \mathcal{P}_3 \) is illustrated on the \( XY \) plane in Fig. \ref{portrait_P4}. It is important to note that the critical point $\mathcal{P}_3$, where $q = 0$ and $\omega = \frac{-1}{3}$, does not depict accelerating expansion. This critical point occurs at $2 m^2 - m \neq 0$ and is stable when $\frac{1}{4} < m < \frac{1}{2}$. The deceleration parameter solely relies on the variable $Z = \frac{\dot{H}}{H^2}$. In this context, when $Z=-1$, it signifies that the contribution from the GB invariant $\mathcal{G} = 24 H^4 (Z+1)$ term vanishes, which could explain why it does not demonstrate the transition phase, as depicted in Fig. \ref{portrait_P4}. The phase portrait shows trajectories spiralling inwards towards \( \mathcal{P}_3 \), indicating that it is a stable spiral. This suggests that small perturbations will cause the system to oscillate while eventually converging to \( \mathcal{P}_3 \). The eigenvalues, consisting of negative real parts, confirm the stable nature of this critical point.

    \item {\bf Critical point \( \mathcal{P}_4 \):}
        The critical point \( \mathcal{P}_4 \) is depicted on the \( XY \) plane in Fig. \ref{portrait_P5}. The critical point $\mathcal{P}_4$ represents the de Sitter solutions with $\Omega_{\text{DE}}=1$, $\Omega_{\text{m}}=0$, $\Omega_{\text{r}}=0$, indicating the current accelerated expansion of the Universe. This de Sitter solution is valid at the critical point $\mathcal{P}_4$ within the parameter range $(-1+2 m) (-1+ 4 m) \neq 0$ and $m \neq 0$. Consequently, the value of $\omega_{\text{eff}}=q=-1$ highlights the significance of this critical point in describing the current dynamics of the Universe. The trajectories show a clear spiral structure converging towards \( \mathcal{P}_4 \), indicating a stable focus. This implies that the system will exhibit damped oscillations as it approaches \( \mathcal{P}_4 \). The stability analysis of \( \mathcal{P}_4 \) supports this observation, showing that the real parts of the eigenvalues are negative, thus confirming stability.
    \end{itemize}
    
        Understanding the stability and characteristics of critical points is essential for gaining insight into the long-term dynamics and behaviour of the specified dynamical system.

    \begin{figure}
    \centering
        \includegraphics[width=85mm]{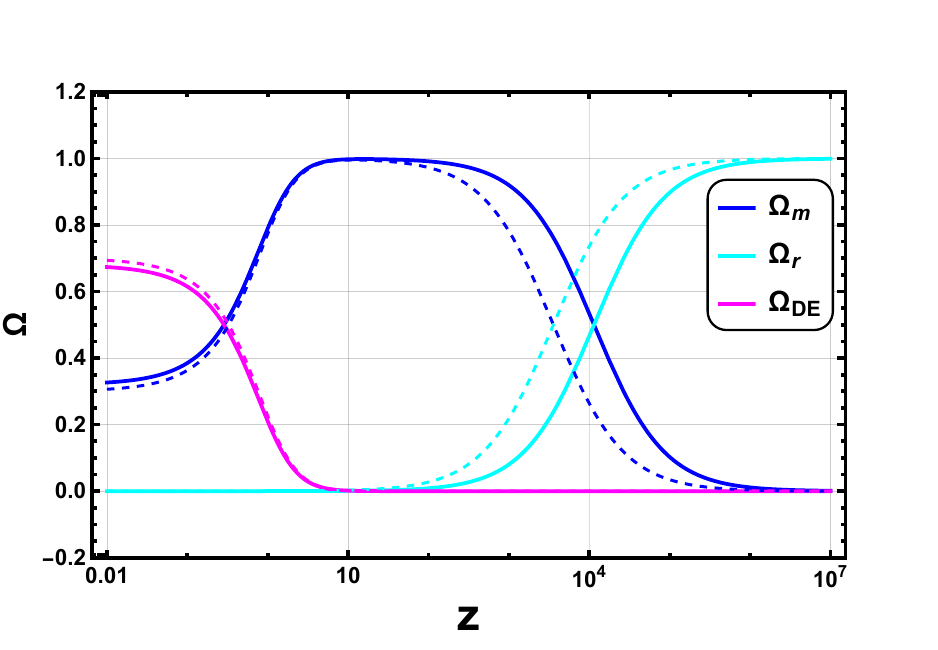}
        \caption{Evolution of the relative energy densities of dark matter $\Omega_{\text{m}}$, radiation $\Omega_{\text{r}}$, and DE $\Omega_{\text{DE}}$. The thick line represents the evolution of the density parameter for the $f(\mathcal{G})$ model, while the dashed line shows the evolution for the $\Lambda$CDM model.} 
    \label{FIG8}
    \end{figure}
    
        In the system described by equations \eqref{Autonomous DS}, it is possible to perform numerical integration with suitable initial conditions to capture the complete cosmological evolution across different epochs. Current measurements of cosmological parameters \citep{Aghanim_2020_641} suggest that the Universe is almost flat. For this specific example, we use the initial conditions $X = 10^{14}$, $Y = -1.2 \times 10^{14}$, $Z = 0.005$, and $V = 8.2 \times 10^{-5}$ and model parameter $m = 0.18$. The behaviour observed aligns with current cosmic observations regarding the evolution of density parameters. By integrating equation \eqref{generaldynamicalsystem} using the summarized initial conditions, we obtain numerical solutions for the density parameters $\Omega_{\text{m}}$, $\Omega_{\text{r}}$, and $\Omega_{\text{DE}}$, as shown in Fig. \ref{FIG8}. These results reveal that the Universe evolves through a radiation-dominated phase at early times $(q = 1)$. Subsequently, it transitioned into a matter-dominated phase with a deceleration parameter $\frac{1}{2}$. Currently, it is moving into an exponentially accelerating epoch with a deceleration parameter of $-1$. The $f(\mathcal{G})$ model represents a cosmological scenario in which the Universe undergoes successive eras of radiation domination, matter domination, and currently, DE domination. Our results indicate that the point where matter and radiation contribute equally is slightly higher than in the $\Lambda$CDM model. The behaviour of the model is consistent with current cosmic observations on the evolution of density parameters. The current densities are approximately $\Omega_{\text{m}} \approx 0.3$, $\Omega_{\text{DE}} \approx 0.7$, and $\Omega_{\text{r}} \approx 10^{-4}$. Similar behaviour in the evolution of density parameters have been noted in the literature (\cite{Granda_2020_80}).

\section{Conclusion} \label{SEC V}
        In this study, we explored the cosmological properties of a specific modified GB gravity model. Initially, we discussed the main features of a gravitational action, which includes a general combination of the Ricci scalar and the GB invariant. Then, assuming a flat FLRW cosmological background, we derived the point-like Lagrangian of the theory and the corresponding equations of motion. The specific function we focused on, $f(\mathcal{G}) = \alpha \sqrt{\beta}(\frac{\mathcal{G}^2}{\beta^2})^m$, approaches GR as the real constant $\alpha$ gets closer to zero. However, our study does not explicitly converge to the cosmological constant case, making it particularly interesting as a potential alternative to the standard $\Lambda$CDM model. This model shows the ability to replicate DE behaviour while avoiding the conceptual issues associated with $\Lambda$. Importantly, we showed that the right-hand sides of the modified Friedmann equations can be understood as effective energy density and pressure resulting from curvature.

        We investigated the cosmological properties of the $f(\mathcal{G})$ model in the presence of matter fields. We assumed non-relativistic pressureless matter and neglected the late-time contribution of the radiation fluid. By numerically solving the first Friedmann equation, we determined the redshift behaviour of the Hubble parameter. We used the $\Lambda$CDM model to establish appropriate initial conditions for $H(z)$ and its derivatives. Subsequently, we utilized the most recent low-redshift observations to compare our theory directly with the model-independent predictions of the cosmic expansion. Specifically, we used a Bayesian analysis with the MCMC method, analysing using the Pantheon$^+$ and CC data sets separately. By assuming uniform prior distributions, we obtained constraints on the free parameters of the model at the 1$\sigma$ and 2$\sigma$ confidence levels. This enabled us to reconstruct the cosmological evolution of the Hubble expansion rate and the total effective EoS parameter. Our analysis indicates that the $f(\mathcal{G})$ model effectively accounts for the current acceleration of the Universe without the need for $\Lambda$. However, upon closer analysis and comparison with the predictions of the standard cosmological scenario, it becomes evident that the $f(\mathcal{G})$ model exhibits significant deviations from $\Lambda$CDM as the redshift increases, demonstrating its inability to describe a standard matter-dominated era. Our analysis identifies a notable discrepancy between the $H_0$ values obtained from the CC sample and the Pantheon data sets, as seen in Fig. \ref{FIG: whisker plot}, highlighting the ongoing $H_0$ tension in cosmology. This result emphasizes the necessity for further research into possible systematic errors or new physics to resolve this issue. Addressing these discrepancies is essential for enhancing our understanding of the expansion rate of the Universe.

        In the second phase of our study, we conducted a dynamical system analysis, focusing on the type of $f(\mathcal{G})$ function under consideration. This analysis has enabled us to assess the global behaviour and stability of the cosmological model. It provided insights into the critical points associated with the model and their characteristics, which could be relevant to observable cosmology and the evolution of the Universe. Table \ref{TABLE-IV} presents the eigenvalues of the critical points along with their corresponding stability conditions. Our findings revealed stable critical points describing the late-time cosmic accelerated phase. This indicates non-standard matter and radiation-dominated eras of the Universe. Interestingly, our results align with the standard quintessence model on $z>0$. We made notable preliminary discoveries regarding the finite phase space of a power-law class of the GB gravity model. In addition, the equations describing the dynamical system for the power law $f(\mathcal{G})$ gravity model are provided in Equations \eqref{Autonomous DS}. Furthermore, Table \ref{TABLE-II} includes critical points, existing conditions, effective EoS, and deceleration parameters for the autonomous system. In contrast, Table \ref{TABLE-III} presents density parameter values for the acceleration phase. In total, we identified four critical points, three being stable ($\mathcal{P}_2$, $\mathcal{P}_3$, $\mathcal{P}_4$) and one unstable ($\mathcal{P}_1$). Finally, Fig. \ref{FIG8} shows the proficiency of the model in depicting the evolution of dark matter, radiation, and DE densities, effectively capturing the transitions through various cosmic epochs and reinforcing the model’s robustness in explaining late-time cosmic phenomena.

\section*{Data Availability}
There are no associated data with this article. No new data were generated or analysed in support of this research.

\section*{Acknowledgments}
SVL would like to express gratitude for the financial support provided by the University Grants Commission (UGC) through the Senior Research Fellowship (UGC Reference No. 191620116597) to carry out the research work. BM acknowledges the support of the Council of Scientific and Industrial Research (CSIR) for the project grant (no. 03/1493/23/EMR II). We sincerely appreciate the esteemed referee and editor for their insightful suggestions, which have greatly enhanced the quality and presentation of our research. 

\bsp	
\label{lastpage}

\end{document}